\documentclass[twocolumn,showpacs,preprintnumbers,amsmath,amssymb,prl]{revtex4}

\usepackage{dcolumn}
\usepackage{bm}
\usepackage{hyperref} 
\usepackage{amsmath}
\usepackage{amsfonts}     
\usepackage{graphics}  
\usepackage{psfrag} 
\usepackage{graphicx}    
\usepackage{epsfig}    
\usepackage{rotating}  
 \usepackage[latin1]{inputenc}
 \usepackage{color}
 \usepackage{rotating}

\definecolor{DarkBlue}{rgb}{0.0,0.0,0.5}
\definecolor{DarkGreen}{rgb}{0.0,0.7,0.0}
\definecolor{LightBlue}{rgb}{0.5,1.0,1.0}
\definecolor{orange}{rgb}{1.0,0.4,0.0}
\definecolor{DarkYellow}{rgb}{0.7,0.7,0}
\definecolor{grey}{rgb}{0.5,0.5,0.5}

\begin{document}


\title{Dual contribution to amplification in the mammalian inner ear}

\author{Tobias Reichenbach}
\author{A.~J.~Hudspeth}

\affiliation{Howard Hughes Medical Institute and Laboratory of Sensory Neuroscience, The Rockefeller University, 1230 York Avenue, New York, NY 10065-6399}


\date{\today}
             
\begin{abstract}
The inner ear achieves a wide dynamic range of responsiveness by  mechanically amplifying weak sounds. The enormous mechanical gain reported for the mammalian cochlea, which exceeds a factor of $4,\!000$, poses a challenge for theory. 
Here we show how such a large gain can result from  an interaction between amplification by low-gain hair bundles and a pressure wave:
hair bundles can amplify both their displacement per locally applied pressure and the pressure wave itself. A recently proposed ratchet mechanism, in which hair-bundle forces do not feed back on the pressure wave, delineates the two effects.
Our  analytical calculations with a WKB approximation agree with numerical solutions.
\end{abstract}

\maketitle

Hearing employs an active process to achieve a remarkable sensitivity, frequency selectivity, and dynamic range~\cite{robles-2001-81,hudspeth-2009-59}. Understanding the cellular basis of the active process in the mammalian cochlea remains a fundamental and controversial topic in contemporary hearing research. Active force production by hair bundles, the sensory organelles of the mechanoreceptive hair cells, underlies the active process in non-mammalian tetrapods~\cite{martin-2001-98,fettiplace-2001-24} and contributes to mammalian hearing~\cite{kennedy-2005-433,chan-2005-8}. The  wide dynamic range of the mammalian cochlea, however, poses a  challenge for active hair-bundle motility as the cochlear amplifier. The cochlea achieves this dynamic range by compressing a large range of input sound intensities into a relatively narrow range of outputs in the form of hair-bundle displacements. This nonlinear compression results from  the active process and reflects its gain. Experimental measurements \emph{in vitro} indicate that  active hair-bundle forces can increase the amplitude of hair-bundle displacements  by a factor of about 10~\cite{martin-2001-98}.  Although the value may be larger \emph{in vivo} and be further increased through coupling of neighboring hair bundles~\cite{dierkes-2008-105}, this low gain falls  orders of magnitude short of the amplification of  $4,\!000$ or more measured in the intact mammalian cochlea~\cite{ruggero-1991-11,robles-2001-81}.
In this Letter we show how  active hair-bundle motility with a low gain can yield a large cochlear gain by interacting with a pressure wave. Although we focus on  hair bundles as the force-producing elements,  our description is more general and the principle of dual amplification applies whenever active forces amplify the basilar-membrane displacement.  Several previous models for the active cochlea have therefore contained this effect implicitly~\cite{neely-1986-79,julicher-2003-90,yoon-2007-122}. Amplification of the pressure wave has not been explicitly stated or quantified previously, however, and the importance of the resulting dual amplification for  cochlear gain has not been recognized.

\begin{figure}[b]
\begin{center}
\includegraphics[width=7cm]{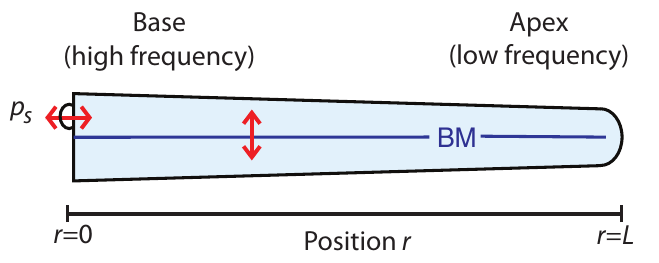}
\end{center}
\vspace*{-0.5cm}
\caption{The mammalian cochlea. Sound displaces the stapes (top left), producing a pressure difference $p_s$ across the basilar membrane (BM) that elicits a traveling wave of pressure difference  and membrane displacement.   \label{fig1}}
\end{figure}

The cochlea consists of two fluid-filled chambers that are separated by the elastic basilar membrane [Fig.~\ref{fig1}]. Sound vibrates the stapes inserted into the oval window at the cochlear base, inducing a pressure difference across the basilar membrane that propagates along the membrane as a traveling wave from  the base towards the apex. 
The physics of the pressure wave and amplification emerges from a one-dimensional model of the cochlea in which the fluid flows in the two interacting chambers are assumed to be constant across a vertical cross-section [Fig.~\ref{fig1}~(a)].  Let $p(r,t)$ denote the pressure difference across the basilar membrane at position $r$ and time $t$ and let $X_\text{BM}(r,t)$ represent the evoked basilar-membrane displacement. The equations of momentum and continuity then yield the wave equation~\cite{lighthill-1981-106}
\begin{equation}
\rho\partial_t^2X_\text{BM}(r,t)+\Lambda\partial_tX_\text{BM}(r,t)=\frac{h}{2}\partial_r^2 p(r,t)\,.
\label{eq:p}
\end{equation}
The phenomenological term including the drag coefficient $\Lambda$  accounts for friction along the boundaries of the cochlear chambers~\cite{Lighthill};  $\rho$ denotes the fluid's density  and $h$ the height of each chamber.

%
%
To solve Eq.~(\ref{eq:p}) we require the dependence of the basilar-membrane displacement $X_\text{BM}(r,t)$ on the pressure difference $p(r,t)$. Consider  stimulation at a single angular frequency $\omega=2\pi f$ such that  $p(r,t)=\tilde{p}(r)e^{i\omega t} + \text{c.c.}$ and $X_\text{BM}(r,t)=\tilde{X}_\text{BM}(r)e^{i\omega t} + \text{c.c.}$ with the Fourier components $\tilde{p}(r)$ and  $\tilde{X}_\text{BM}(r)$ and with ``c.c.'' denoting the complex conjugate.  In the passive cochlea $\tilde{X}_\text{BM}(r)$ depends linearly on $\tilde{p}(r)$: 
\begin{equation}
i\omega Z^\text{pass}(r,\omega)\tilde{X}_\text{BM}(r)=A_\text{BM}\tilde{p}(r)\,,
\end{equation}
in which $Z^\text{pass}(r,\omega)$ represents the local passive impedance  of a transverse strip of the basilar membrane of area $A_\text{BM}$. The \emph{local sensitivity} $|\tilde{X}_\text{BM}(r)/\tilde{p}(r)|$, the magnitude of the  basilar-membrane response to the local pressure, is given by $|A_\text{BM}/[\omega Z^\text{pass}(r,\omega)]|$ and is thus independent of $\tilde{p}$.  In a normal cochlea, though, the active process increases the basilar-membrane displacement and  introduces a nonlinearity. Because the active process counters viscous damping, it can poise each segment of the basilar membrane near an oscillatory instability at the local characteristic frequency~\cite{magnasco-2003-90,julicher-2003-90}. In the vicinity of the resulting Hopf bifurcation the basilar membrane's response to varied pressures is inherently nonlinear. The nonlinear response, however, arises over only  a limited range of pressures. Large pressures yield the passive linear response, for they are not amplified. Small pressures also yield a linear response but with an increased gain;  linearity  arises in this instance because the system does not operate exactly at the bifurcation and because of noise~\cite{julicher-2009-29}. For these small pressures we may write
\begin{equation}
i\omega Z^\text{act}(r,\omega)\tilde{X}_\text{BM}(r)=A_\text{BM}\tilde{p}(r)
\end{equation}
with the local active impedance $Z^\text{act}(r,\omega)$. The magnitude of the ratio between the linear active  and the linear passive responses, $|Z^\text{pass}(r,\omega)/Z^\text{act}(r,\omega)|$, represents the local gain. 

Eq.~(\ref{eq:p}) may be solved through the WKB approximation 
when the basilar-membrane displacement depends linearly on $\tilde{p}(r)$~\cite{lighthill-1981-106}. Assume  $i\omega Z(r,\omega) \tilde{X}_\text{BM}(r)=A_\text{BM}\tilde{p}(r)$, in which  $\tilde{Z}(r,\omega)$ represents either the passive or the active basilar-membrane impedance.
 For a pressure $\tilde{p}(r_0)=p_s$ applied at the stapes and for the case of a forward-travelling wave, the ansatz 
\begin{equation}
\tilde{p}(r)=a(r)e^{-i\omega b(r)}
\end{equation}
yields, to orders $\omega^2$ and $\omega$ respectively,
\begin{equation}
b(r)=\int_0^r dr'\frac{1}{c(r')}\,\,\text{and} \,\,
a(r)=p_s\sqrt{\frac{c(r)}{c(0)}}e^{-\Lambda b(r)/(2\rho)}\,,
\label{eq:ab}
\end{equation}
with the wave's velocity
\begin{equation}
c(r)=\sqrt{\frac{i\omega h Z(r,\omega)}{2\rho A_\text{BM}}}\,.
\label{eq:c}
\end{equation}
The magnitude of the pressure  follows as 
\begin{equation}
|\tilde{p}(r)|=|a(r)|e^{\omega \text{Im}[b(r)]}\,.
\label{eq:p_magn}
\end{equation}

At each position along the basilar membrane, the local impedance defines a resonant frequency. Assume that $Z(r,\omega)$ results from  mass $m(r)$,  viscous damping $\lambda(r)$, and stiffness $K(r)$:
\begin{equation}
Z(r,\omega)=i\omega m(r)+\lambda(r) -iK(r)/\omega\,.
\label{eq:Z}
\end{equation}
The mass and stiffness yield a resonant frequency $\omega_0(r)=\sqrt{K(r)/m(r)}$ at which the imaginary part of $Z(r,\omega)$ vanishes and changes sign: $\text{Im}[Z(r,\omega)]<0$ for 
$\omega<\omega_0(r)$ but  $\text{Im}[Z(r,\omega)]>0$ for $\omega>\omega_0(r)$.    It follows from Eqs.~(\ref{eq:ab}),(\ref{eq:c}) that, in the absence of basilar-membrane friction,  the pressure wave can travel along the basilar membrane as long as $\omega<\omega_0(r)$ and thus up to the resonant position $r_0$ defined by $\omega=\omega_0(r_0)$. Indeed, basal to  $r_0$  the wave velocity $c(r)$ is real because $\text{Im}[Z(r,\omega)]<0$. Upon approaching $r_0$, $\text{Im}[Z(r,\omega)]$ vanishes and therefore  $c(r)$ also tends to zero. Apical to the resonant position, where $\text{Im}[Z(r,\omega)]>0$, the wave velocity $c(r)$ becomes imaginary.  The amplitude of the pressure wave thus declines upon approaching the resonant position, for it is proportional to $\sqrt{c(r)}$ [Eq.~(\ref{eq:ab})]. The basilar-membrane displacement $X_\text{BM}(r)$ varies in proportion to $[c(r)]^{-3/2}$, however,  and therefore diverges at the resonant position $r_0$.
Viscous forces dominate the basilar-membrane impedance at  $r_0$ and yield a finite wave velocity as well as a finite displacement. 

%
%
\begin{figure}[t]
\begin{center}
\includegraphics[width=8.5cm]{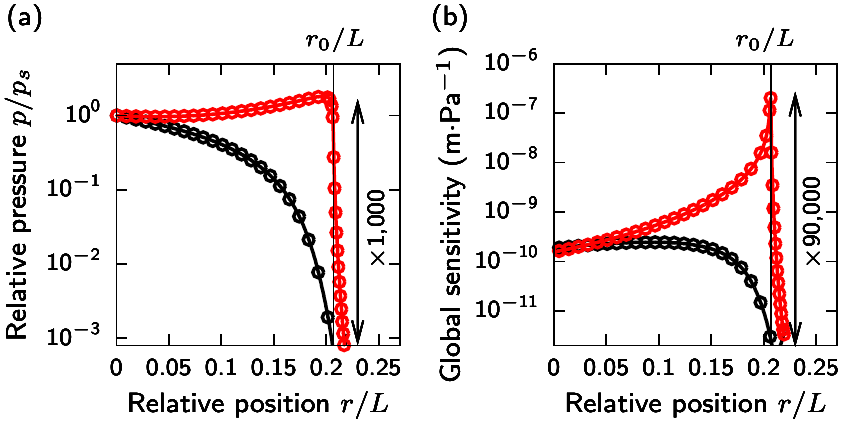}
\end{center}
\vspace*{-0.5cm}
\caption{
Cochlear pressure and global sensitivity of the basilar membrane  from numerical solution of Eq.~(\ref{eq:p}) (lines) and from the WKB approximation (circles) relative to $p_s$ for a frequency $f=8$~kHz. (a),   Pressure (red, active; black, passive). (b), Global sensitivity  (red, active; black, passive). The gain in global sensitivity exceeds the gain in pressure by the factor $|\lambda^\text{act}/\lambda^\text{pass}|=90$. The results in this figure and the following have been obtained using the two-mass model for the organ of Corti and parameters from Ref.~\cite{reichenbach-2010}; $\rho=10^3$~kg$\cdot$m$^{-3}$. 
  \label{WKB}}
\end{figure}
Amplification counteracts viscous damping in two ways.  First, it increases the basilar membrane's local sensitivity. At the resonant position the basilar-membrane impedance includes only  the viscous contribution, such that the displacement varies in inverse proportion to the damping coefficient: $\tilde{X}_\text{BM}(r_0)=-iA_\text{BM}\tilde{p}(r_0)/[\omega\lambda(r_0)]$.  We assume amplification to reduce the damping coefficient in the basilar-membrane impedance [Eq.~(\ref{eq:Z})] from the passive value $\lambda^\text{pass}(r)$ to a smaller value $\lambda^\text{act}(r)$ and consequently to yield a gain in basilar-membrane displacement, and thus in hair-bundle displacement, of $|\lambda^\text{pass}/\lambda^\text{act}|$. Experiments on the dynamics of hair bundles demonstrate this effect~\cite{martin-2001-98}: a small force applied directly to a hair bundle elicits an \emph{in vitro}  displacement that is about a factor of $10$ greater for an active than for a passive bundle~\cite{martin-2001-98}.

The second effect of amplification is to enhance the amplitude of the pressure wave itself. The term $\text{Im}[b(r)]$ in the  contribution $e^{\omega \text{Im}[b(r)]}$ to the pressure magnitude [Eq.~(\ref{eq:p_magn})] represents the imaginary part of the integrated inverse wave speed and results from damping [Eq.~(\ref{eq:ab})]. Because  $\text{Im}[c^{-1}(r)]$ is approximately proportional to $-\lambda(r)$ away from the resonant position, damping occurs at a strength proportional to the integrated viscosity. A reduced damping coefficient $\lambda^\text{act}(r)$ basal to the resonant position therefore  diminishes damping and augments the pressure wave, yielding a gain of $\exp\{{\omega\text{Im}[b^\text{act}(r_0)]-\omega\text{Im}[b^\text{pass}(r_0)]}\}\approx\exp\{-\omega\text{Im}[b^\text{pass}(r_0)]\}$.   Because this increase represents the \emph{cumulative} reduction in damping, its magnitude can  significantly exceed the gain in local sensitivity that follows from the reduced \emph{local} damping alone.  

What  is the  magnitude of the gain in pressure amplitude? Experimental measurements on the travelling wave's phase, $\omega\text{Re}[b(r)]$, indicate that the wave undergoes about two cycles while traveling from the stapes to its resonant position~\cite{robles-2001-81}: $\omega \text{Re}[b(r_0)]\approx 4\pi$. The imaginary part of $c^{-1}(r)$ is smaller than the real part distant from the resonant position, but comparable in its vicinity.  The integrated imaginary part of $c^{-1}(r)$, $\text{Im}[b(r_0)]$ is thus smaller than but, for the passive case, presumably of the same order of magnitude as its integrated real part, $\text{Re}[b(r_0)]$.  Because $e^{2\pi}$ is  about $500$, active hair-bundle motility can enormously enhance the amplitude of the pressure wave near the resonant position [Fig.~\ref{WKB}(a)]. Measurements of the intracochlear pressure near the basilar membrane confirm the amplification of the pressure amplitude~\cite{olson-1999-402}.

The \emph{global sensitivity} $|\tilde{X}_\text{BM}(r,\omega)/p_s|$, the magnitude of the basilar-membrane movement in response to the pressure at the stapes, is subject both to the gain in local sensitivity and to pressure-wave amplification. The net gain for this dual amplification follows as the product of the two individual gains and can exceed $10,\!000$ [Fig.~\ref{WKB}(b)].  For realistic parameter values, numerical solution of the wave equation~(\ref{eq:p}) validates the  WKB approximation and shows that the amplitude of the  amplified pressure wave exceeds the passive value by a factor of about $1,\!000$ at the resonant position [Fig.~\ref{WKB} (a)]. The local basilar-membrane sensitivity experiences an additional gain near $90$, resulting in an overall gain of about  $90,\!000$ [Fig.~\ref{WKB} (b)]. 
 
Damping of the pressure wave also results from friction through the term $e^{-\Lambda \text{Re}[b(r)]/(2\rho)}$ in the pressure amplitude [Eq.~(\ref{eq:ab})]. Because active hair-bundle motility presumably does not change the imaginary part of the impedance $Z(r,\omega)$, which includes the inertial and elastic contributions, it should not significantly alter $ \text{Re}[b(r)]$ and therefore not counter this type of friction.

Amplification causes a compressive nonlinearity in the hair bundle's response to varied sound-pressure levels.
The dominant nonlinearity presumably results from the nonlinear dependence of the open probability $P$ of ion channels in the hair bundle on its deflection $X_\text{HB}$.
The mechanotransduction channels are situated at the tips of the hair bundle's stereocilia and are connected by filamentous  tip links to neighboring stereocilia. Deflection of the hair bundle in the excitatory direction  pulls transduction channels open, with the open probability following a Boltzmann distribution:
\begin{equation}
P(X_\text{HB})=\left[1+e^{-B X_\text{HB}}\right]^{-1}\,.
\label{eq:P}
\end{equation}
The coefficient $B$ encodes the energy release due to channel opening. For small and large values of $X_\text{HB}$,  $P(X_\text{HB})$ is asymptotically linear. For intermediate hair-bundle displacements, however, a  nonlinearity emerges that is predominantly cubic because outer hair cells operate at a symmetry point around a resting open probability $P(X_\text{HB}=0)=0.5$. 

 %
%
\begin{figure}[t]
\begin{center}
\includegraphics[width=8.7cm]{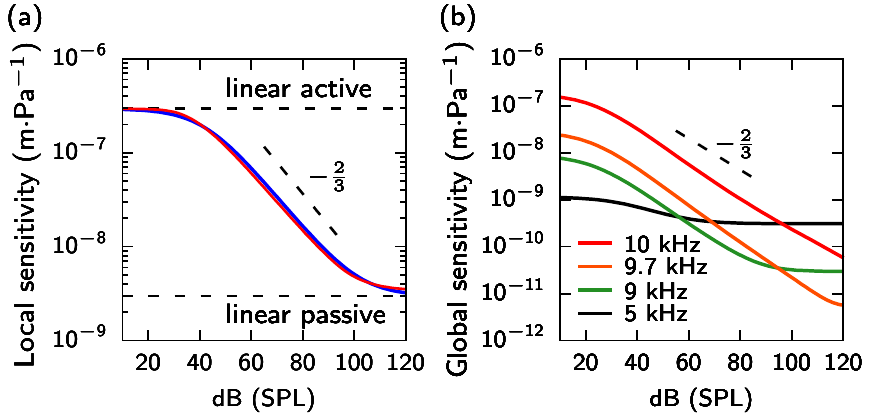}
 \end{center}
 \vspace*{-0.3cm}
\caption{Nonlinearities at the $10$~kHz resonant position. (a), Local sensitivity of hair-bundle displacement. Numerical solutions with the open probability given by Eq.~(\ref{eq:P}) (red)  agree with the approximation of Eq.~(\ref{eq:Xnonlin}) (blue). (b), Global sensitivity  for different sound frequencies. The response becomes linear as  the frequency deviates from the resonant frequency.  \label{cochlea_nonlin} }
\end{figure} 
For variations in  the locally applied pressure  the hair bundle's displacement exhibits a cubic nonlinearity as well. 
The release of tension in tip links during  channel opening, which is proportional to $P-P_0$, produces a hair-bundle force  $F_\text{HB}=-F_\text{TL}(P-P_0)$ with a coefficient $F_\text{TL}$~\cite{kennedy-2005-433}. Because of  feedback from  molecular motors, this force  can counter viscous damping and poise the bundle near a Hopf bifurcation~\cite{choe-1998-95,vilfan-2003-85,tinevez-2007-11}. The hair-bundle response  can be  approximated by
\begin{equation}
\tilde{X}_\text{HB}=e\left(\frac{f+\tilde{p}}{g+\tilde{p}}\right)^{2/3}\tilde{p}\,.
\label{eq:Xnonlin}
\end{equation}
The coefficients $e$ and $f$ follow from  the linear active response for small pressure differences, $e(f/g)^{2/3}\tilde{p}$,  and the linear passive response for large pressure differences, $e\tilde{p}$. The constant $g$ determines
the location of the intermediate nonlinear regime in which the cubic nonlinearity $\tilde{X}_\text{HB} = ef^{2/3}\tilde{p}^{1/3}$ emerges. The agreement with a numerical solution is excellent [Fig.~\ref{cochlea_nonlin}~(a),(b)]~\footnote{The parameter values are as in~\cite{reichenbach-2010}; $F_\text{TL}$  follows a logarithmic map from $20$~nN  (base) to $50$~pN (apex)~\cite{kennedy-2005-433},  $B=0.1~\text{m}^{-1}$, and $g=0.5$~Pa.}. The cubic nonlinearity introduces a slope of $-2/3$ in the local sensitivity to varying pressure [Fig.~\ref{cochlea_nonlin}~(a)].

Hair-bundle displacement is related to basilar-membrane displacement. Substituting $\tilde{X}_\text{BM}$ by $\tilde{X}_\text{HB}$ in Eq.~(\ref{eq:p}) through a model for the organ of Corti's micromechanics~\cite{reichenbach-2010} and subsequently substituting $\tilde{X}_\text{HB}$ by $\tilde{p}$ through Eq.~(\ref{eq:Xnonlin}) we arrive at a wave equation for $\tilde{p}$  that we can solve numerically.
Because of dual amplification the resulting compressive nonlinearity in basilar-membrane and hair-bundle  motion in response to a pressure $p_s$ at the stapes  extends over a significantly broader range of sound intensities than the nonlinearity per locally applied pressure [Fig.~\ref{cochlea_nonlin}~(a),(b)]. Amplification  of the pressure wave follows from a decreased damping term $e^{\omega\text{Im}[b(r)]}$, so the resulting nonlinearity reflects properties of the traveling wave and the basilar membrane  basal to the characteristic point. We still find in our numerics an approximately cubic nonlinearity [Fig.~\ref{cochlea_nonlin}~(b)]. 

Dual amplification arises when hair-bundle force feeds back onto the basilar membrane and amplifies its motion, thus enhancing the pressure wave. However, this feedback can be avoided: hair-bundle motion could decouple from basilar-membrane motion, omitting its amplification even if the reverse coupling were maintained. This intriguing type of unidirectional mechanical coupling can arise from electromotility, the ability of the outer hair cell's body to elongate and contract in response to electrical stimulation~\cite{reichenbach-2010}. In this ``ratchet mechanism,'' forces acting on the basilar membrane elicit hair-bundle motion whereas the reverse does not hold. The ratchet mechanism may underly hearing in the mammalian cochlea at  frequencies below $1-2$~kHz~\cite{reichenbach-2010}.

 %
%
 \begin{figure}[t]
\begin{center}
\includegraphics[width=8.7cm]{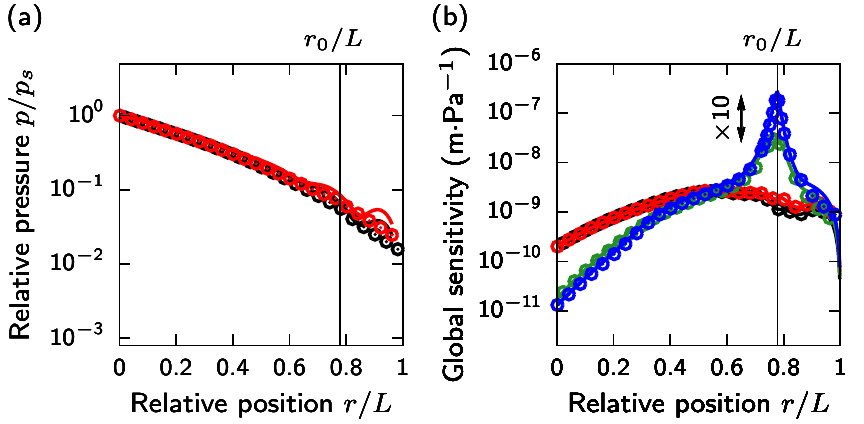}
\end{center}
\vspace*{-0.3cm}
\caption{
Pressure and sensitivity for amplification through the ratchet mechanism; $f=200$ Hz. The data result from numerical solution of Eq.~(\ref{eq:p}) (lines) and from the WKB approximation (circles). (a) Pressure (red, active; black, passive). (b), Gobal sensitivity of the hair bundles (blue, active; green, passive) and of the basilar membrane (red, active; black, passive).  \label{WKB_2}}
\end{figure}
The ratchet mechanism separates amplification of hair-bundle motion per local pressure difference from amplification of the pressure wave. Because in the ratchet mechanism active hair-bundle forces do not amplify basilar-membrane motion, they do not enhance the pressure wave [Fig.~\ref{WKB_2} (a)]. The net hair-bundle gain is low, about $10$, for it reflects solely the enhanced local sensitivity of hair bundles [Fig.~\ref{WKB_2} (b)]. The compressive nonlinearity associated with this low gain encompasses a much smaller dynamic range than for dual amplification as observed experimentally in the apical half of the cochlea~\cite{robles-2001-81}.

We have  demonstrated that active hair-bundle motility can enhance hair-bundle displacement  in two ways. First, active hair-bundle force increases a bundle's local sensitivity, with a plausible gain of $10-100$. Second, hair-bundle forces can feed back onto the basilar membrane and therefore enhance the amplitude of the pressure wave itself, yielding a gain of $100$ and greater. The overall cochlear gain,  the product of these two components, can exceed $10,\!000$.

This work was supported by grant DC000241 from the National Institutes of Health.
T.~R. acknowledges support from a fellowship of the Alexander von Humboldt
Foundation. A.~J.~H. is an Investigator of Howard Hughes Medical Institute.



\end{document}